\RequirePackage{ifpdf}
\ifpdf 
\documentclass[pdftex]{sigma}
\else
\documentclass{sigma}
\fi

\begin{document}

\allowdisplaybreaks

\renewcommand{\PaperNumber}{081}

\FirstPageHeading

\renewcommand{\thefootnote}{$\star$}

\ShortArticleName{QFT on the Fuzzy Sphere}

\ArticleName{Quantum Field Theory in a Non-Commutative Space: \\
Theoretical Predictions and Numerical Results\\ on the Fuzzy
Sphere\footnote{This paper is a contribution to the Proceedings of
the O'Raifeartaigh Symposium on Non-Perturbative and Symmetry
Methods in Field Theory
 (June 22--24, 2006, Budapest, Hungary).
The full collection is available at
\href{http://www.emis.de/journals/SIGMA/LOR2006.html}{http://www.emis.de/journals/SIGMA/LOR2006.html}}}

\Author{Marco PANERO} 
\AuthorNameForHeading{M. Panero}

\Address{School of Theoretical Physics, Dublin Institute for Advanced Studies,\\
10 Burlington Road, Dublin 4, Ireland}
\Email{\href{mailto:klimcik@iml.univ-mrs.fr}{panero@stp.dias.ie}}
\URLaddress{\href{http://www.stp.dias.ie/~panero/}{http://www.stp.dias.ie/\~{}panero/}}

\ArticleDates{Received September 29, 2006, in f\/inal form
November 10, 2006; Published online November 17, 2006}

\Abstract{We review some recent progress in quantum f\/ield theory
in non-commutative space, focusing onto the fuzzy sphere as a
non-perturbative regularisation scheme. We f\/irst introduce the
basic formalism, and discuss the limits corresponding to
dif\/ferent commutative or non-commutative spaces. We present some
of the theories which have been investigated in this framework,
with a particular attention to the scalar model. Then we comment
on the results recently obtained from Monte Carlo simulations, and
show a preview of new numerical data, which are consistent with
the expected transition between two phases characterised by the
topology of the support of a matrix eigenvalue distribution.}

\Keywords{non-commutative geometry; matrix models; non-perturbative effects; phase transitions}

\Classification{58B34; 81R60}

\section{Introduction}

The study of quantum f\/ield theory in non-commutative
spaces~\cite{Doplicher:1994tu, Landi:1997sh, Douglas:2001ba,
Szabo:2006wx} is based on strong physical motivations, since this
formalism is relevant to the description of the quantum Hall
ef\/fect~\cite{Karabali:2004xq}, to certain aspects of string
theory~\cite{Connes:1997cr, Douglas:1997fm, Alekseev:1999bs,
Seiberg:1999vs, Kar:1999tu, Kar:2000pb}, as well as to a possible
formulation for a quantum theory of
gravity~\cite{Chamseddine:1992yx}.

The most extensively studied non-commutative space is probably the
Groenewold--Moyal $\mathbb{R}^n_\theta$ space, def\/ined as the
algebra of functions ${\cal A}_\theta(\mathbb{R}^n)$ generated by
the coordinates $x^\mu$ ($\mu \in \{ 1, 2, \dots , n \}$) with the
commutation relations:
\begin{equation}
\label{moyalnc} \left[x^\mu, x^\nu \right]_* = i \theta^{\mu\nu} ,
\end{equation}
where, for simplicity, $\theta^{\mu\nu}$ can be taken to be
independent of $x$, and $*$ denotes the Groenewold--Moyal
star-product. For functions $f$, $g$ in ${\cal
A}_\theta(\mathbb{R} ^n)$, the latter is def\/ined to act
as~\cite{Groenewold:1946kp, Moyal:1949sk, Szabo:2001kg}:
\begin{equation*}
f(x)*g(x)= f(x) e^{\frac{i}{2} \overleftarrow{\partial}_\mu
\theta^{\mu\nu}\overrightarrow{\partial}_\nu}g(x)  . \nonumber
\end{equation*}
A number of dramatic implications stem from
equation~(\ref{moyalnc}), which have been studied in detail in the
literature~\cite{Doplicher:1994zv, Doplicher:2001qt, Filk:1996dm,
Chen:2001qg, Jain:2003xs, Chaichian:2004za, Szabo:2001kg,
Minwalla:1999px, Hinchliffe:2002km, Gubser:2000cd,
Castorina:2003zv,  Langmann:2002cc, Grosse:2004yu,
Rivasseau:2005bh, Govindarajan:2006vh, Grosse:2005ig,
Grosse:2006qv, Grosse:2006tc}. In particular, some issues which
are of strong relevance from the physical point of view include:
\begin{itemize}
\itemsep=0pt \item The ``ultra-violet/infra-red (UV/IR) mixing''
phenomenon: The ef\/fective action describing QFT in a
Groenewold--Moyal space diverges when the components of the
external momentum along the non-commutative directions vanish.
This ef\/fect arises from the integration of high-energy modes in
non-planar loop diagrams, and is related to the non-locality
properties of quantum f\/ield theory in a non-commutative space.

\item Renormalisability: One of the potentially  most dangerous
consequences of the UV/IR mixing phenomenon is the fact that it
threatens the possibility to renormalise quantum f\/ield theories
def\/ined in a non-commutative setting. However, the proof of
perturbative renormalisability at all orders has been eventually
worked out adding an oscillator term to the potential of the
scalar model~\cite{Grosse:2004yu, Rivasseau:2005bh, Grosse:2005ig,
Grosse:2006qv, Grosse:2006tc}.
\end{itemize}

\renewcommand{\thefootnote}{\arabic{footnote}}
\setcounter{footnote}{0}

The theoretical predictions for these models can be compared with
non-perturbative results obtained from numerical simulations.  In
fact, it is possible to def\/ine a well-suited lattice formulation
of the non-commutative theory as a unitary matrix
model~\cite{Ambjorn:1999ts}, which possesses a continuum limit
characterised by f\/inite space volume and f\/inite
non-commutativity. This formulation allowed to investigate
numerically various aspects of these models~\cite{Ambjorn:2000nb,
Ambjorn:2000cs, Azuma:2004zq, Bietenholz:2002ch,
Bietenholz:2002vj, Ambjorn:2002nj, Bietenholz:2004xs,
Bietenholz:2005iz, Bietenholz:2006cz}.

A dif\/ferent class of non-commutative spaces is given by fuzzy
spaces; the basic idea underlying their construction is to use a
f\/inite-dimensional matrix algebra to approximate the
inf\/inite-dimensional algebra of functions on a manifold. This
can be done on even-dimensional co-adjoint orbits of Lie groups,
which are symplectic manifolds~\cite{Madore:1991bw,
Alexanian:2001qj, Balachandran:2001dd, Hammou:2001cc,
Medina:2002pc, Vaidya:2003ew, Dolan:2003kq, Balachandran:2005ew,
Sheikh-Jabbari:2006bj}, like the two-sphere $S^2$ and the
$\mathbb{C} P^n$ complex projective spaces.

In the following discussion, we shall concentrate our attention
onto the fuzzy two-sphere $S^2_F$~\cite{Madore:1991bw}; it
provides the simplest example of a fuzzy space, and depends on two
parameters: the dimensionful, real and positive, radius $R$, and a
dimensionless, integer-valued cut-of\/f, which is associated with
the value of the maximum angular momentum $l_{\max}$. In
particular, the ordinary, commutative algebra of functions on the
sphere is replaced by a non-commutative counterpart, obtained
truncating the expansion of functions in the basis of irreducible
representations of the $su(2)$ algebra to a maximum angular
momentum $l_{\max}$. This geometrical construction of $S^2_F$
of\/fers a natural representation for functions on the sphere in
terms of matrices belonging to Mat$_N (\mathbb{C})$, with
$N=l_{\max}+1$\footnote{This construction can be easily obtained
by taking the tensor product of two vector representations of
angular momentum $\frac{l_{\max}}{2}$.}.

The fuzzy approach explicitly maintains the symmetries of the
original manifold for every value of $N$, and it accommodates
various topological features in a natural way --
see~\cite{Balachandran:2005ew} and references therein for details.
It provides a well-def\/ined, non-perturbative regularisation
scheme for quantum f\/ield theory, and -- at least in principle --
can be considered as a potential alternative to the lattice
discretisation. The representation of f\/ields on the fuzzy sphere
in terms of f\/inite-dimensional matrices also opens up the
possibility to study interacting theories via numerical
simulations: this has motivated various recent
works~\cite{Martin:2004un, GarciaFlores:2005xc, Medina:2005su,
Azuma:2004yg, Anagnostopoulos:2005cy, Azuma:2005pm,
O'Connor:2006wv, Panero:2006bx}.

Indeed, the fuzzy models reveal a rich and complex structure,
encoding a number of interesting properties, which show up even
for the simplest theories -- e.g.: for a scalar f\/ield with
quartic interactions.

The aim of this paper consists in presenting an overview on some
of these aspects, comparing theoretical expectations and results
of Monte Carlo simulations, and highlighting their physical
interpretation in dif\/ferent limits.

The sections below are organised according to the following
structure.  In Section~\ref{formalismsect}, we f\/irst introduce
the general formalism for the construction of the fuzzy sphere,
and discuss its dif\/ferent geometrical limits; then we def\/ine
the description of QFT on the fuzzy sphere, and present the
theoretical predictions which are expected in dif\/ferent limits.
Next, the theoretical predictions are compared with
non-perturbative results obtained from numerical simulations in
Section~\ref{comparisonsect}. Finally, in
Section~\ref{conclusionsect} we discuss the relevance of these
results for the program to use the the fuzzy approach as a
regularisation scheme for quantum f\/ield theory in commutative or
in non-commutative spaces, and conclude with some brief remarks.

\section{General formalism and theoretical predictions}\label{formalismsect}

The construction of a fuzzy space is based on the quantisation  of
the algebra of functions def\/ined on the corresponding
(commutative) manifold. For the two-sphere, the construction can
be summarised as it follows~\cite{Grosse:1995ar}: The
inf\/inite-dimensional algebra of polynomials in the $\{ x_i
\}_{i=1,2,3}$ commutative coordinates on the two-dimensional
sphere $x_i x_i = R^2$ embedded in $\mathbb{R}^3$ is replaced by
the non-commutative algebra generated by $\{ \hat{x}_i
\}_{i=1,2,3}$ operators, which obey the (rescaled) commutation
relations of the $su(2)$ algebra:
\begin{equation}
\left [ \hat{x}_i,  \hat{x}_j \right] = \frac{2R}{\sqrt{N^2-1}} i
\epsilon_{ijk} \hat{x}_k , \qquad \mbox{with:}\quad  \sum_{i=1}^3
\hat{x}_i^2 = R^2  .  \label{commutationrelations}
\end{equation}

On the fuzzy sphere, the counterpart of functions def\/ined on the
commutative sphere are matrices belonging to Mat$_N (\mathbb{C})$;
a natural mapping can be def\/ined truncating the basis of
harmonic functions on $S^2$ to a maximum value for the angular
momentum $l_{\max}$, and associating the harmonic functions with
total and third-component angular momentum quantum numbers~$l$
and~$m$ to the $\{ \hat{Y}_{l,m} \}$ polarisation tensors
satisfying:
\begin{gather*}
\big[ L_i , \big[ L_i, \hat{Y}_{l,m}\big] \big] = l(l+1)\hat{Y}_{l,m} , \qquad 
\big[ L_3, \hat{Y}_{l,m}\big]= m \hat{Y}_{l,m}  , 
\end{gather*}
where the $\{ L_i \}_{i=1,2,3}$ are the generators of the $su(2)$
algebra in the $N \times N$ matrix representation. This mapping
choice, however, is not unique -- a convenient alternative mapping
being the one def\/ined in terms of coherent states.

Integrals of functions on the commutative sphere are replaced by
the matrix trace operation: given any $\Phi$ and $\Psi$ elements
in Mat$_N (\mathbb{C})$, an inner product on $S^2_F$ is built as:
\begin{equation*}
\langle \Phi, \Psi \rangle = \frac{4\pi R^2}{N} \mbox{tr} \big(
\Phi^\dagger \Psi \big) . \nonumber
\end{equation*}

The vectors describing derivations on the commutative sphere are
represented through the adjoint action of the $\{ L_i
\}_{i=1,2,3}$ generators:
\begin{equation*}
[ L_i , \Phi ]  .
\end{equation*}

It is important to discuss what is the meaning of the
``commutative limit'' in this context: The matrix algebra
underlying the fuzzy space is truly non-commutative for any $N$;
however, the right-hand side of
equation~(\ref{commutationrelations})  vanishes in the $N
\rightarrow \infty$ limit, keeping $R$ f\/ixed. Therefore, given a
smooth function def\/ined on the commutative sphere, whose
expansion in the spherical harmonics' basis involves
non-negligible coef\/f\/icients up to a certain angular momentum
only, the ef\/fect of this discretisation will be vanishing in the
large-$N$ limit at f\/ixed $R$.

So far, the interpretation of this limit is only concerned with
the geometric properties of the manifold -- not with the physical
properties of a quantum f\/ield theory def\/ined in this setting.

Indeed, the quantum features of a physical f\/ield theory spoil
the correspondence between the f\/ixed-$R$, large-$N$ limit of the
(naive formulation of the) theory on the fuzzy sphere, and the
corresponding model on the commutative manifold~\cite{Chu:2001xi}.

This can be seen in a perturbative calculation for the (real)
scalar model with quartic interactions; for this model, the
Euclidean action on the fuzzy sphere can be written in the
following form:
\begin{equation}\label{action}
S \left( \Phi\right)= \frac{4\pi}{N} \mbox{tr}  \left(\frac{1}{2}
\left[ L_i, \Phi \right]^\dagger \left[ L_i, \Phi \right] +
\frac{\mu^2}{2} \Phi^2 + \frac{g}{4!} \Phi^4 \right)
\end{equation}
(where $\Phi \in \mbox{Mat}_N (\mathbb{C})$ is  taken to be
Hermitian). At one-loop order in a perturbative expansion, the
non-commutative nature of the model manifests itself in the
comparison between the contributions to the
one-particle-irreducible two-point function from planar
($I_{\mbox{\tiny{P}}}$) and non-planar ($I_{\mbox{\tiny{NP}}}$)
diagrams -- they dif\/fer by a quantity which is f\/inite,
non-vanishing, and which has a smooth dependence on $\frac{1}{N}$:
\begin{equation}
\label{planarnonplanardifference} I_{\mbox{\tiny{NP}}} =
I_{\mbox{\tiny{P}}} - \frac{2}{N^2 -1} \sum_{j=0}^{N-1}
\frac{j(j+1)(2j+1)}{j(j+1)+\mu^2} .
\end{equation}
When the $N \rightarrow \infty$, f\/ixed-$R$, limit is taken, this
dif\/ference between $I_{\mbox{\tiny{NP}}}$ and
$I_{\mbox{\tiny{P}}}$ has an impact on the one-loop ef\/fective
action, as it induces a weakly non-local, $l$-dependent,
deformation of the dispersion relation on the fuzzy sphere. This
ef\/fect has no analogue in the commutative setting, and is
therefore called the ``non-commutative anomaly''.

In order to recover the correct model in the commutative limit,
one should redef\/ine the interaction term in  in
equation~(\ref{action}) via a normal-ordering
prescription~\cite{Dolan:2001gn}, cancelling the undesired
momentum-dependent quadratic terms in the ef\/fective action;
alternatively, one could include rotationally symmetric higher
derivative terms in the matrix action~\cite{briandenjoe}.

On the contrary, studying the model def\/ined by the (naive)
matrix action in equation~(\ref{action}), one expects to observe
deviations with respect to the quantum theory on the commutative
sphere. Although the derivation of the non-commutative anomaly
in~\cite{Chu:2001xi} was worked out for the $\mu^2 > 0$ case, one
may expect that a similar phenomenon occurs when $\mu^2 < 0$, too.

For QFT on the commutative sphere, the latter regime is
particularly interesting, as it corresponds to a situation in
which, classically, the $\Phi \rightarrow - \Phi$ symmetry of the
model is spontaneously broken. The situation from the quantum
point of view is dif\/ferent, however, because -- due to the
f\/inite size of the system -- tunneling events characterised by a
f\/inite Euclidean action connect the classical vacua, and the
true quantum ground state is actually unique for any f\/inite
value of the parameters. Therefore, strictly speaking, one does
not expect phase transitions on the commutative sphere with
f\/ixed, f\/inite radius; this, however, might no longer be the
case for the $N \rightarrow \infty$ limit of the fuzzy model, if
``anomalous'' ef\/fects take place.

For the quantum theory on the commutative sphere, it is also
interesting  to note that the tunneling phenomenon between the
(uniform) classical vacua can be mediated by typical quantum
events which break the $SO(3)$ symmetry -- for instance, through a
$p$-wave f\/ield conf\/iguration (with axial symmetry only). A
signature of these events would show up through non-vanishing
expectation values for the modes associated with non-vanishing
momenta. In the commutative setting, this phenomenon is possible
because of the f\/inite size of the system; on the contrary, one
would not expect a non-uniform breaking symmetry pattern in the
inf\/inite commutative plane.

Another important implication of
equation~(\ref{planarnonplanardifference}) is that the
contribution of non-perturbative diagrams to the two-point
function \emph{is not} divergent for vanishing values of the
momentum. This has to be compared and contrasted with the
non-commutative plane setting, where the UV/IR mixing phenomenon
occurs. Yet, the divergence which is encountered in the the
Groenewold--Moyal space is exactly recovered, once a dif\/ferent
scaling limit of the fuzzy sphere is considered: this is discussed
below.

A local approximation to the non-commutative $\mathbb{R}^2_\theta$
plane can be obtained, in the following double scaling limit of
the fuzzy sphere:
\begin{equation}
\label{doublescalinglimit} N \rightarrow \infty , \qquad R
\rightarrow \infty , \qquad \mbox{with:} \quad \theta
=\frac{2R^2}{N} \ \ \mbox{f\/ixed.}
\end{equation}
This can be seen considering a stereographic projection map  from
the sphere to the $\left( \hat{y}_1, \hat{y}_2 \right)$ plane:
\begin{gather*}
\hat{y}_+=\hat{y}_1 + i \hat{y}_2 = 2 R \hat{x}_+ \left( R -
\hat{x}_3 \right)^{-1} ,\qquad \hat{y}_-=\hat{y}_1 - i \hat{y}_2 =
2 R \left( R - \hat{x}_3 \right)^{-1} \hat{x}_-  .
\end{gather*}
In the limit~(\ref{doublescalinglimit}), $\hat{y}_1$ and
$\hat{y}_2$ satisfy:
\begin{equation*}
\left[ \hat{y}_1, \hat{y}_2\right] = - i \theta
\end{equation*}
and the non-commutative anomaly induces a logarithmic infrared
divergence proportional to $\log \left( p \theta \right)$ in the
non-planar contribution to the 1PI two-point function. This
divergence is exactly the same which is found in the perturbative
calculation for the quantum theory def\/ined in the
Groenewold--Moyal plane~\cite{Minwalla:1999px}, therefore in this
case the (double scaling) large-$N$ limit of the fuzzy sphere
reproduces the expected result, and the fuzzy space can be used as
a regularisation for non-commutative spaces. In principle, the
construction can be generalised to higher dimensional cases, using
fuzzy $\mathbb{C} P^n$ spaces, or direct products of fuzzy
spheres.

The regularisation of Euclidean scalar f\/ield theory in
even-dimensional, non-commutative $\mathbb{R}^n_\theta$ spaces by
means of fuzzy manifolds was used in~\cite{Steinacker:2005wj}.
There, the properties of the model were investigated
non-perturbatively in a matrix representation for the f\/ield, and
a critical line corresponding to a phase transition in the
associated eigenvalue distribution was derived in the large-matrix
limit.

This transition is a change in the topology of the support of the
eigenvalue distribution: for values of $\mu^2$ above a critical
value (``single-cut phase''), the eigenvalue distribution
$\rho(\phi)$ has a~connected support of f\/inite width, centred in
zero, and would reduce to Wigner's semi-circle law in the
non-interacting case:
\begin{equation}
\label{eigenvaluedistribution} \rho(\phi)= \frac{1}{2\pi} \left[
g_{\mbox{\tiny{eff}}} \left( \phi^2 + \frac{1}{2} \right) +
\mu_{\mbox{\tiny{eff}}}^2 \right] \sqrt{1-\phi^2}, \qquad
\mbox{with: } g_{\mbox{\tiny{eff}}} =\frac{4}{3} \left( 4 -
\mu_{\mbox{\tiny{eff}}}^2 \right)  ,
\end{equation}
where $g_{\mbox{\tiny{eff}}}$ and $ \mu_{\mbox{\tiny{eff}}}^2 $
depend on (the renormalised values of) $g$ and $\mu^2$.

By contrast, for values of $\mu^2$ which are negative and large in
modulus, there exist two symmetric, disconnected regions of the
real axis with non-vanishing probability for the eigenvalues
(``two-cut phase''). In particular, the probability density for
the zero eigenvalue vanishes when $\mu^2$ is decreased down to the
critical point, and for lower $\mu^2$ values a gap opens up in the
distribution. This phenomenon (which would not take place for a
model in ordinary commutative space) is related to the UV/IR
mixing ef\/fect; it is predicted on the basis of a matrix model
description, which relies on the dominance of planar diagrams over
the non-planar ones, and can be interpreted as a manifestation of
the ``striped phase''~\cite{Gubser:2000cd, Castorina:2003zv}.

\begin{figure}[t]
\centerline{\begin{minipage}[t]{75mm} \centering
\includegraphics[width=75mm]{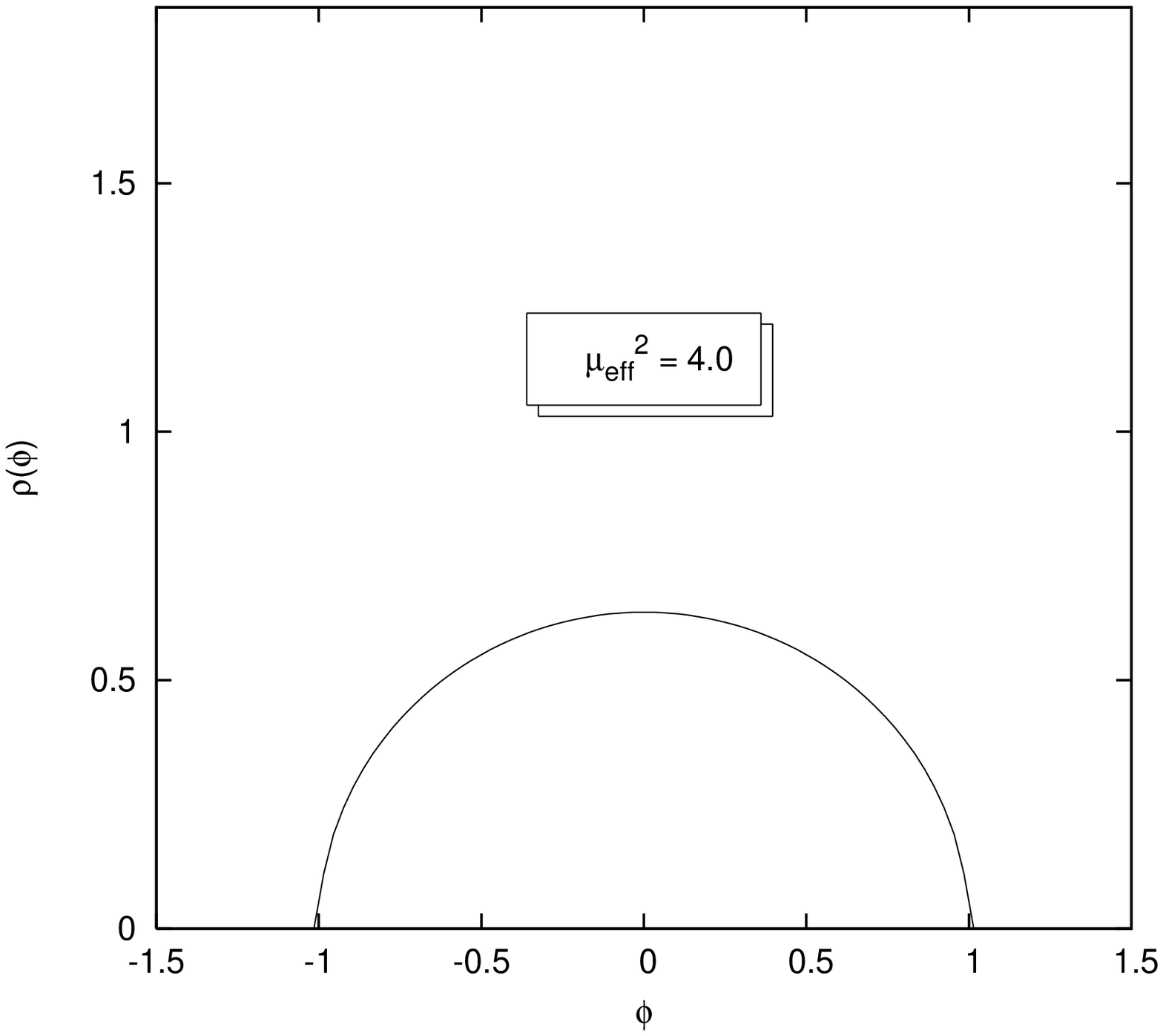}
\caption{Eigenvalue distribution corresponding to Wigner's
semi-circle law: this is obtained setting
$\mu_{\mbox{\tiny{eff}}}^2=4$ in
equation~(\ref{eigenvaluedistribution}).} \label{semicircle_fig}
\end{minipage}\hfill
\begin{minipage}[t]{75mm}
\centering
\includegraphics[width=75mm]{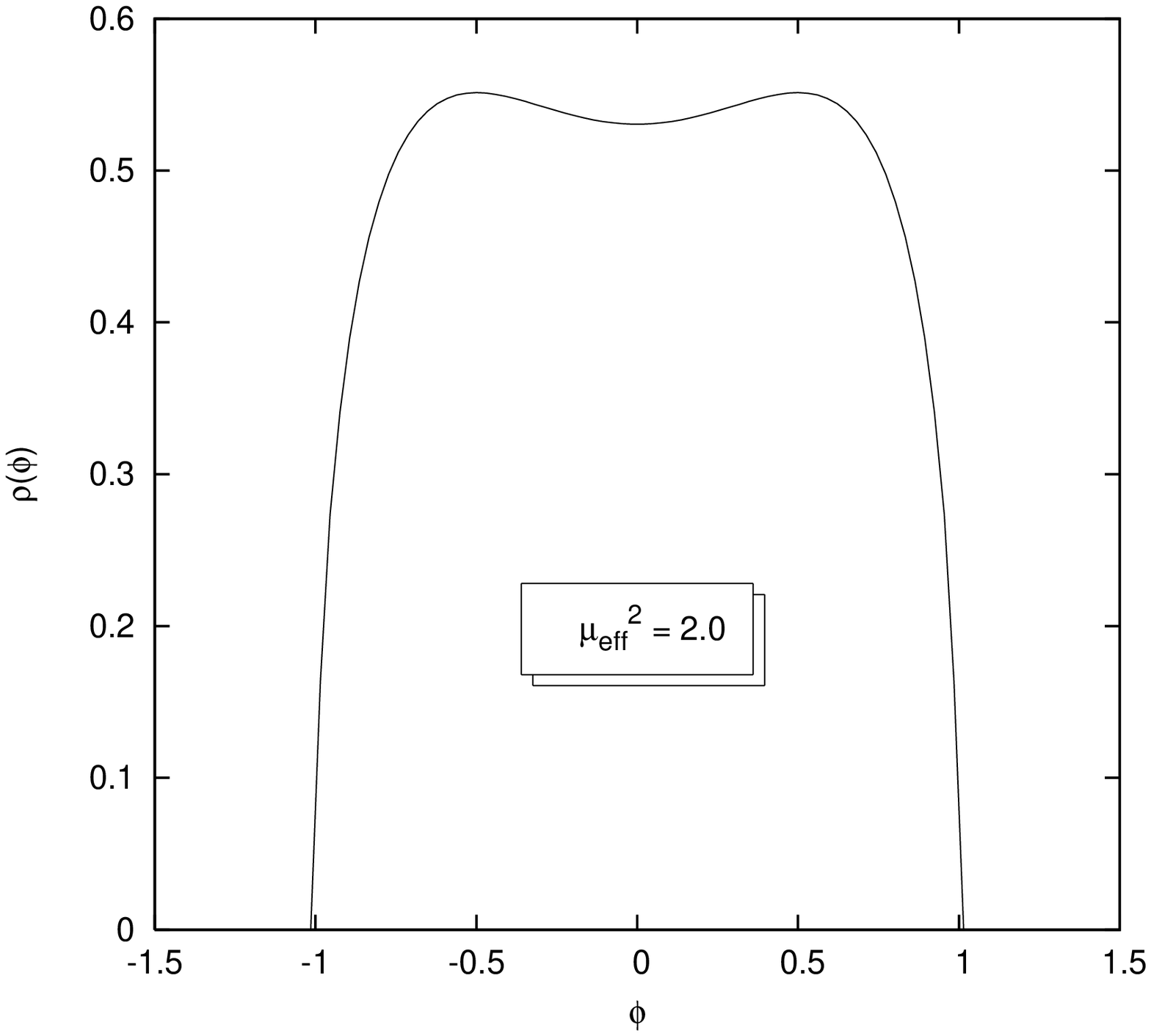}
\caption{Typical prof\/ile of the eigenvalue distri\-bution in the
``one-cut phase''.} \label{onecut_fig}
\end{minipage}}
\vspace{-3mm}
\end{figure}

\begin{figure}[t]
\centerline{\begin{minipage}[t]{75mm} \centering
\includegraphics[width=75mm]{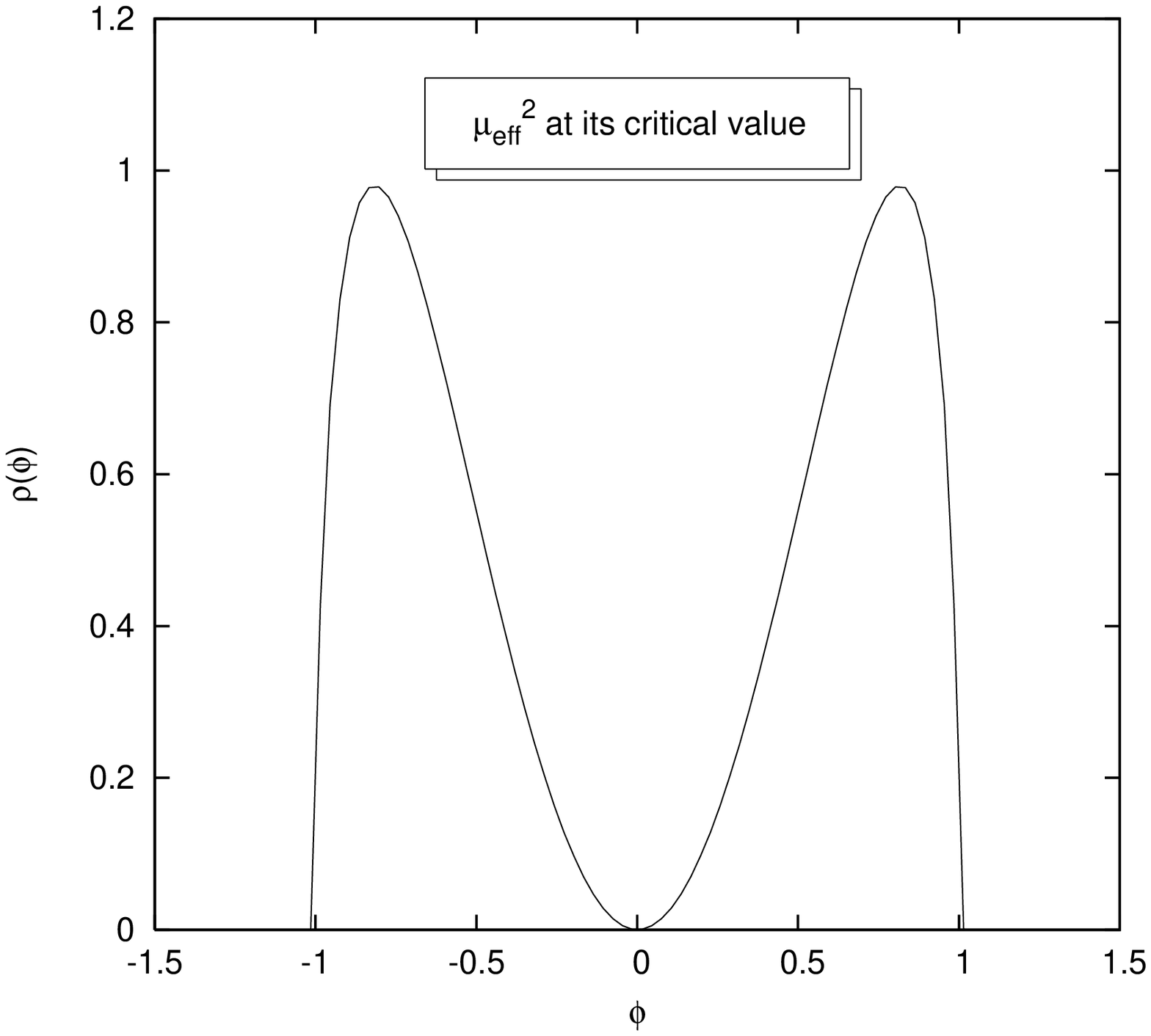}
\caption{At the transition point, the probabili\-ty to observe the
zero eigenvalue becomes vani\-shing.} \label{transition_fig}
\end{minipage}\hfill
\begin{minipage}[t]{75mm}
\centering
\includegraphics[width=75mm]{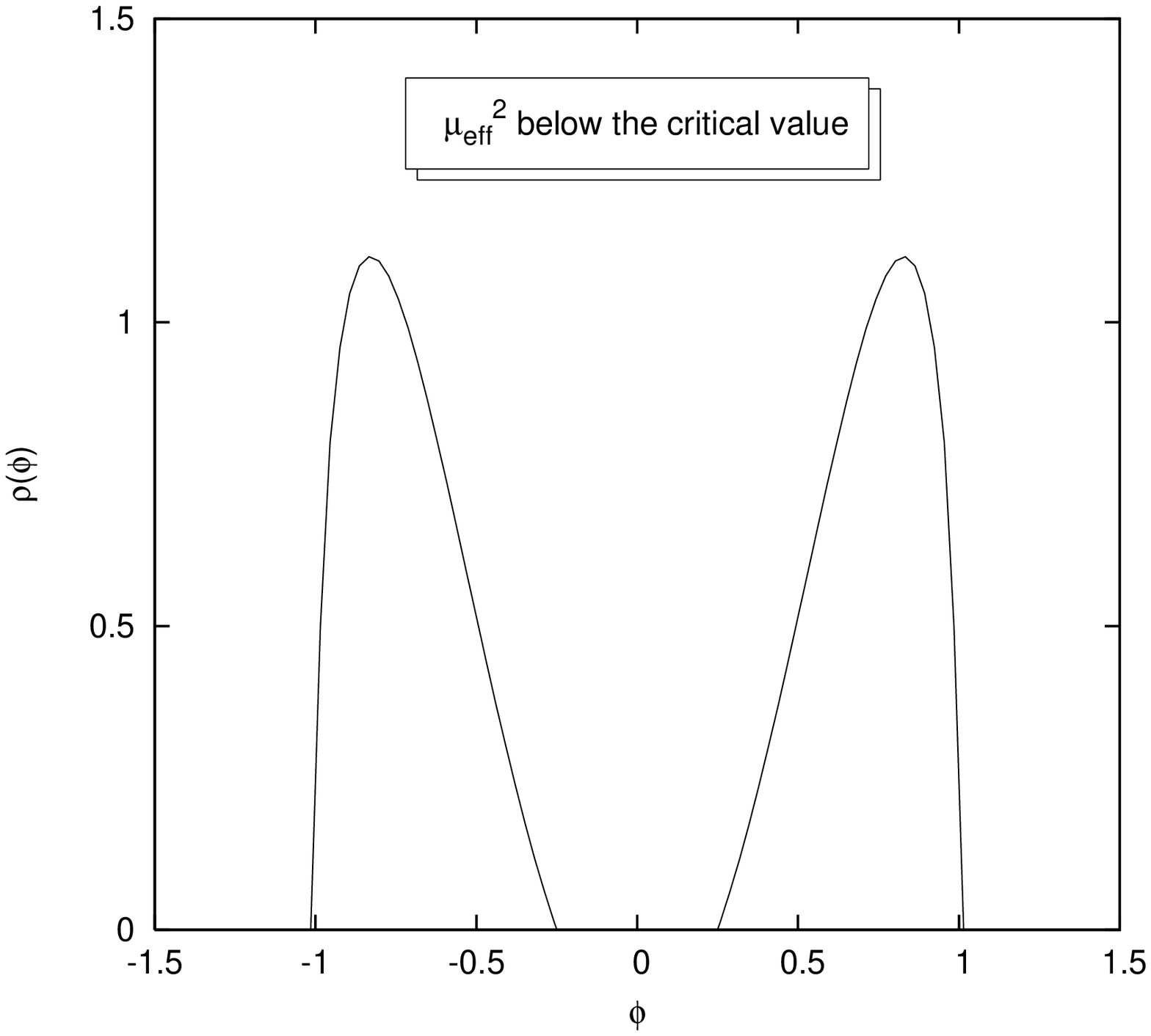}
\caption{In the ``two-cut phase'', $\rho(\phi)$ is zero in a
f\/inite neighbourhood of the origin, and shows two disconnected
peaks.} \label{twocuts_fig}
\end{minipage}}
\vspace{-3mm}
\end{figure}

Figs.~\ref{semicircle_fig} to~\ref{twocuts_fig} show $\rho(\phi)$
for dif\/ferent cases. Fig.~\ref{semicircle_fig} corresponds to
the pure Wigner's semi-circle law; next, Fig.~\ref{onecut_fig}
displays the eigenvalue density prof\/ile in the ``one-cut''
regime. When $\mu_{\mbox{\tiny{eff}}}^2$ is decreased down to a
critical value, $\rho(0)$ vanishes -- Fig.~\ref{transition_fig} --
and for lower values of $\mu_{\mbox{\tiny{eff}}}^2$ the support of
the distribution breaks down into two disconnected components
(Fig.~\ref{twocuts_fig}).

The arguments underlying the derivation
in~\cite{Steinacker:2005wj} are expected to be stronger in the
four-dimensional case than in the two-dimensional setting, because
in $D=2$ the dominance of the planar diagrams over the non-planar
ones is weaker; however, even in $D=2$, one can still look for a
signature of the expected ef\/fects -- and, as it will be
discussed in Section~\ref{comparisonsect}, the agreement with the
numerical results appears quite remarkable.

A dif\/ferent model was studied in~\cite{Medina:2005su}: it is a
scalar f\/ield theory in three-dimensional space, in which two
coordinates are regularised using a fuzzy sphere, while the third
one (labelled as $t$) is considered as purely commutative, and is
regularised via a conventional lattice discretisation:
\begin{gather}
S \left( \Phi\right) = \frac{4  \pi R^2 \Delta t}{N}
\sum_{t=1}^{N_t} \mbox{tr} \Bigg\{ \frac{1}{2R^2}
\Phi\left(t\right)
\left[ L_i,  \left[ L_i,  \Phi\left(t\right) \right] \right]\nonumber\\
 \phantom{S \left( \Phi\right) =}{} +
\frac{1}{2} \left[ \frac{\Phi(t+\Delta t) -\Phi(t)}{\Delta t}
\right]^2 + \frac{m^2}{2} \Phi^2(t) +\frac{\lambda}{4} \Phi^4(t)
\Bigg\}  .\label{julietamodel}
\end{gather}
Also in this case, dif\/ferent limits can be investigated, and one
may expect a general pattern qualitatively similar to the
lower-dimensional case discussed above.

On the other hand, a fuzzy space description for gauge theory was
studied in some papers, including~\cite{Klimcik:1997mg,
Carow-Watamura:1998jn, Alekseev:2000fd, Hashimoto:2001xy,
Kimura:2001uk, Steinacker:2003sd, Iso:2003fp, Kimura:2004hf,
Azuma:2004qe, Grosse:2004wm, Castro-Villarreal:2004vh,
Behr:2005wp, O'Connor:2006wv}.

Supersymmetric versions of these models have been proposed, too.
In particular, a supersymmetric generalisation of the fuzzy sphere
was discussed in~\cite{Grosse:1995pr, GrosseReiter,
Balachandran:2002jf, Balachandran:2005qa}, essentially based on
the $osp(2|1)$ algebra. Other works where the interplay between
supersymmetry and the features of fuzzy models were discussed
include~\cite{Iso:2001mg, Iso:2003zb, Hasebe:2004yp,
Anagnostopoulos:2005cy, Iso:2003fp, Kurkcuoglu:2003ke,
Imai:2003jb, Imai:2003vr}.

\section{Focus on the scalar model in two dimensions:\\
 a comparison with numerical results}\label{comparisonsect}

The models described in Section~\ref{formalismsect} can be
quantised in the path integral approach, def\/ining the
expectation values of generic observables
$\mathcal{O}=\mathcal{O}(c_{l,m})$ as:
\begin{equation}\label{expectationvalues}
\langle \mathcal{O} \rangle = \frac{\int \Pi_{l,m} d c_{l,m}
\mathcal{O}(c_{l,m})e^{-S}}{\int \Pi_{l,m} d c_{l,m} e^{-S}}  ,
\end{equation}
where the $c_{l,m}$ are the components of the $\Phi$ matrix in a
given basis (e.g.: in the polarisation tensor one), and can be
interpreted as the dynamical degrees of freedom of the matrix
model.

The right-hand side of equation~(\ref{expectationvalues}) can be
evaluated perturbatively, or estimated numerically from Monte
Carlo simulations, in an approach similar to standard lattice
f\/ield theory computations, namely: via averages over a sample of
(independent) conf\/igurations with statistical weight
proportional to $\exp \left[ -S\left( \Phi\right) \right]$.

In this section, we summarise the present status of numerical
results for the models discussed above, with a particular
attention to the scalar theory on the two-dimensional fuzzy
sphere. Our aim is to discuss the general physical interpretation
of these results; therefore, we encourage the readers to refer to
the papers mentioned, for precise details about the results, and
for most of the technical aspects associated with the simulations
and with the measurement of dif\/ferent observables.

The real scalar model with quartic interactions on the fuzzy
sphere was studied numerically f\/irst in~\cite{Martin:2004un},
where it was pointed out that the model exhibits three dif\/ferent
phases: a disorder phase (with $\Phi$ f\/luctuating around zero),
a uniform order phase (in which $\Phi$ typically f\/luctuates
around either of the two classical minima of the potential), and
the new, intermediate, non-uniform order phase. In particular, the
critical line separating the disorder phase from the non-uniform
order phase was identif\/ied, and the critical value for $\mu^2$
was found to scale like:
\begin{equation*}
\mu^2_{\mbox{\tiny{crit}}} \simeq -\frac{0.56 N}{R}  .
\end{equation*}
In this work, it was also pointed out that the non-uniform order
phase  can be interpreted in terms of a pure potential
model~\cite{Shimamune:1981qf, Bleher:2002ys}, which can be
obtained from equation~(\ref{action}) neglecting the kinetic term.
The argument goes as follows: The minima of the potential term in
equation~(\ref{action}) belong to orbits with representatives of
the form:
\begin{equation*}
\Phi_{\mbox{\tiny{min}}}^{(p)} = \sqrt{-\frac{6\mu^2}{g}}
 \left( - \mbox{1 \kern-.59em {\rm l}}_p \oplus \mbox{1 \kern-.59em {\rm l}}_{N-p}  \right)  ,
 \qquad \mbox{with:} \quad p \in \{ 0, 1, 2, \dots , N\}  .
\end{equation*}
Unless the kinetic term lifts the degeneracy associated to the $ p
\in \{ 1, 2, \dots , N-1 \}$  orbits, they are dominating -- by
virtue of their larger volume in the phase space -- and imply
non-maximal contributions to the expectation value of the matrix
trace modulus\footnote{It is worth noting, however, that the
dominating conf\/igurations in the model are actually those
characterised by a smooth eigenvalue distribution.}.

The results display a well-behaved collapse of data when $R^2$ is
properly scaled, and are (qualitatively and approximately also
quantitatively) consistent with the predictions
in~\cite{Steinacker:2005wj}: the non-perturbative data seem to lie
in between the theoretical expectation of~\cite{Steinacker:2005wj}
(which, in $D=2$, is only an approximate one), and the pure
potential model limit, in which the kinetic term is completely
neglected.

The results of this study were later extended
in~\cite{GarciaFlores:2005xc}, where large statistics were
accumulated, in order to pin down the other critical line which
appears in the phase diagram, namely: the one separating the
non-uniform order from the uniform order phase. There, the phase
diagram of the model was obtained, in terms of well-suited scaling
combinations of the $\mu^2$ and $g$ parameters with powers of $N$.
In particular, the two dif\/ferent transition lines meet at a
triple point. In the large-$N$ limit, the critical line between
the disorder and the non-uniform order regimes is expressed by a
relation of the form: $g \propto \left( \mu^2_{\mbox{\tiny{crit}}}
\right)^2 $, whereas the transition curve separating the uniform
order from the non-uniform order regimes appears to be compatible
with a straight line (remarkably, the collapse of data obtained in
terms of the simple scaling laws of $g$ and $\mu^2$ works very
well, even down to matrix sizes which are as small as $N=2$).

In the same paper, the behaviour of the specif\/ic heat per degree
of freedom was worked out, and compared with the theoretical
prediction from the large-$N$ limit of the pure potential model.
The data for increasing matrix sizes appear to approach the
expected behaviour, which would imply a discontinuity of the
f\/irst kind in the derivative of the specif\/ic heat p.d.o.f.
with respect to $\left( \frac{R\mu^2}{N\sqrt{g}} \right)$.

A recent work~\cite{Panero:2006bx} has ref\/ined the numerical
analysis of this model, by means of a new method, which reduces
the correlation problems af\/fecting simpler simulation
algorithms. In the $\mu^2>0$ regime, this allowed to observe an
indirect signature of the non-commutative anomaly, as a distortion
of the dispersion relation for non-vanishing momenta. In the
opposite regime, the high-precision numerical results allowed to
highlight a more detailed interpretation of the matrix
conf\/igurations associated with $l>0$ modes. In particular, the
latter are not necessarily incompatible with a commutative
setting: non-uniform conf\/igurations would also contribute the
path integral of QFT on the f\/inite-radius commutative sphere,
and mediate the tunneling among the classical potential minima.
Therefore, their signature in results of simulations in the
large-$N$ limit at f\/ixed $R$ is plausible, and not directly
related to the UV/IR mixing.

The situation, of course, is dif\/ferent, once one investigates
the double scaling limit; in fact, as we discussed above, the
meaning of the latter consists in ``removing'' the regularisation
cut-of\/f $N$, while keeping the non-commutativity parameter
f\/ixed. In this case, the  UV/IR mixing \emph{is} expected (as
the theory tends towards an approximation of the non-commutative
plane), and should be responsible for ef\/fects which do not have
a commutative counterpart. This description in the double scaling
limit is also compatible with the picture of the pure potential
model~\cite{Shimamune:1981qf, Bleher:2002ys}, as it was discussed
in~\cite{Martin:2004un, GarciaFlores:2005xc}.

\begin{figure}[t]
\centerline{\begin{minipage}[t]{75mm} \centering
\includegraphics[width=75mm]{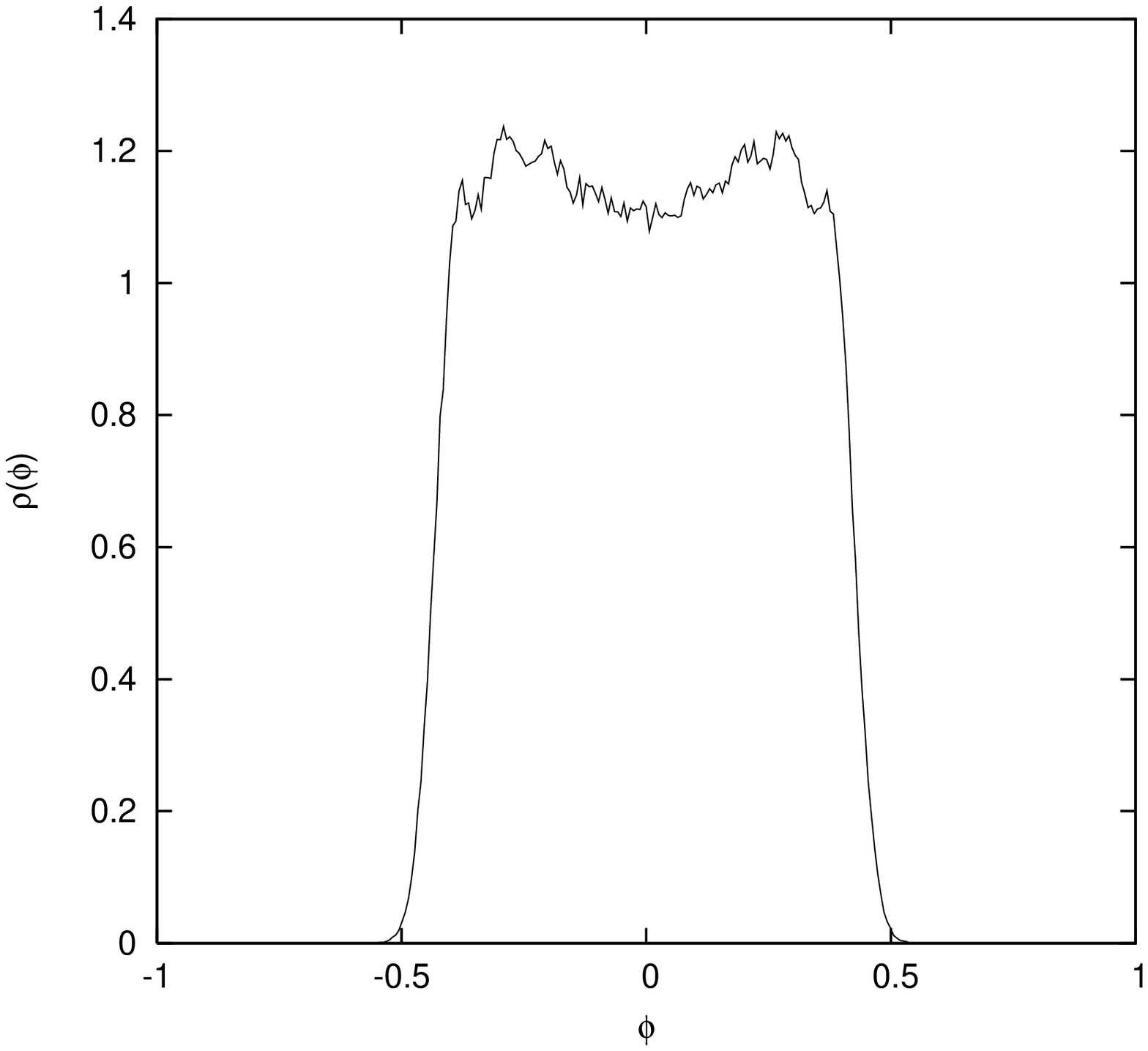}
\caption{Numerical results for the eigenvalue distribution in the
``one-cut'' regime, from data obtained in preliminary,
low-statistics runs with $N=9$, $R=1$, $g=972$ and $\mu^2=0$.}
\label{numerical_onecut_fig}
\end{minipage}\hfill
\begin{minipage}[t]{75mm} \centering
\includegraphics[width=75mm]{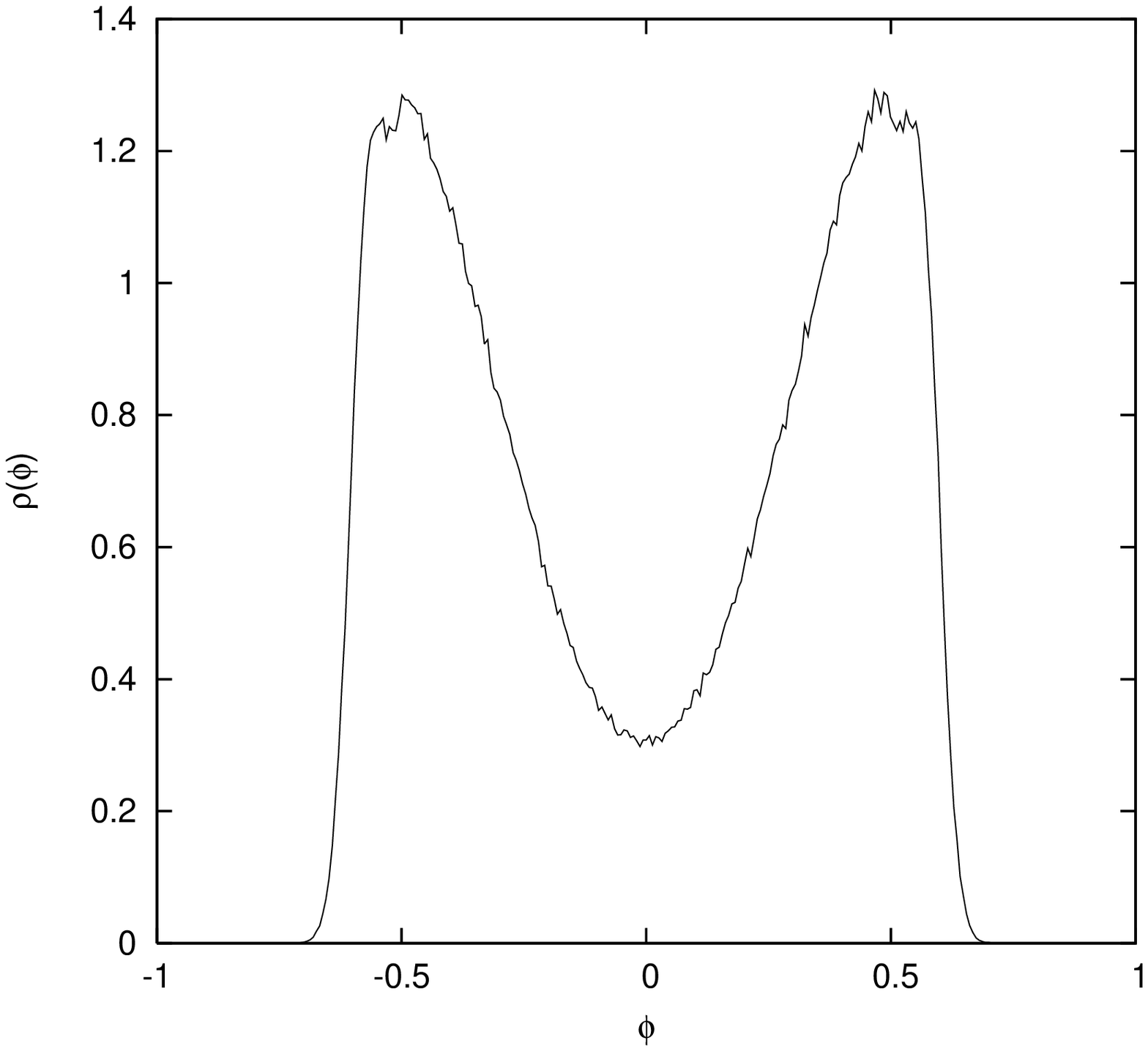}
\caption{As $\mu^2$ is reduced and the transition point is
approached, the eigenvalue density around $\phi=0$ decreases. The
data shown correspond to $\mu^2=-74.25$, at the same $N$, $R$ and
$g$ values as in Fig.~\ref{numerical_onecut_fig}.}
\label{numerical_almost_transition_fig}
\end{minipage}}
\vspace{-3mm}
\end{figure}

The best probe to detect the non-commutative ef\/fects in the
regime where the transition from the disorder to the non-uniform
order phase takes place is probably the density of the matrix
eigenvalues. It is possible that dif\/ferent observables may be
more appropriate for other regions in the parameter space, but at
the moment a general theoretical understanding of the whole phase
diagram of the model is not consolidated yet. Here, we shall only
consider the transition from the disorder to the non-uniform order
phase: for this region, a clear theoretical prediction
exists~\cite{Steinacker:2005wj}, which is formulated precisely in
terms of the eigenvalue distribution. Technically, the measurement
of the $\rho(\phi)$ distribution appears to be simpler than other
observables which could allow to locate the phase transition, and
an algorithm like the one used in~\cite{Panero:2006bx}, which is
highly ef\/f\/icient in the exploration of the phase space of the
model, allows to obtain signif\/icant results within a limited
amount of CPU time.

In fact, the study of the eigenvalue distribution in the
double-scaling limit is currently in progress, and the complete
results will probably be published in an (upcoming) second version
of that paper. The preliminary results obtained so far look very
encouraging, showing a behaviour which conf\/irms the predictions
in~\cite{Steinacker:2005wj}.

\begin{figure}[t]
\centerline{\begin{minipage}[t]{75mm} \centering
\includegraphics[width=75mm]{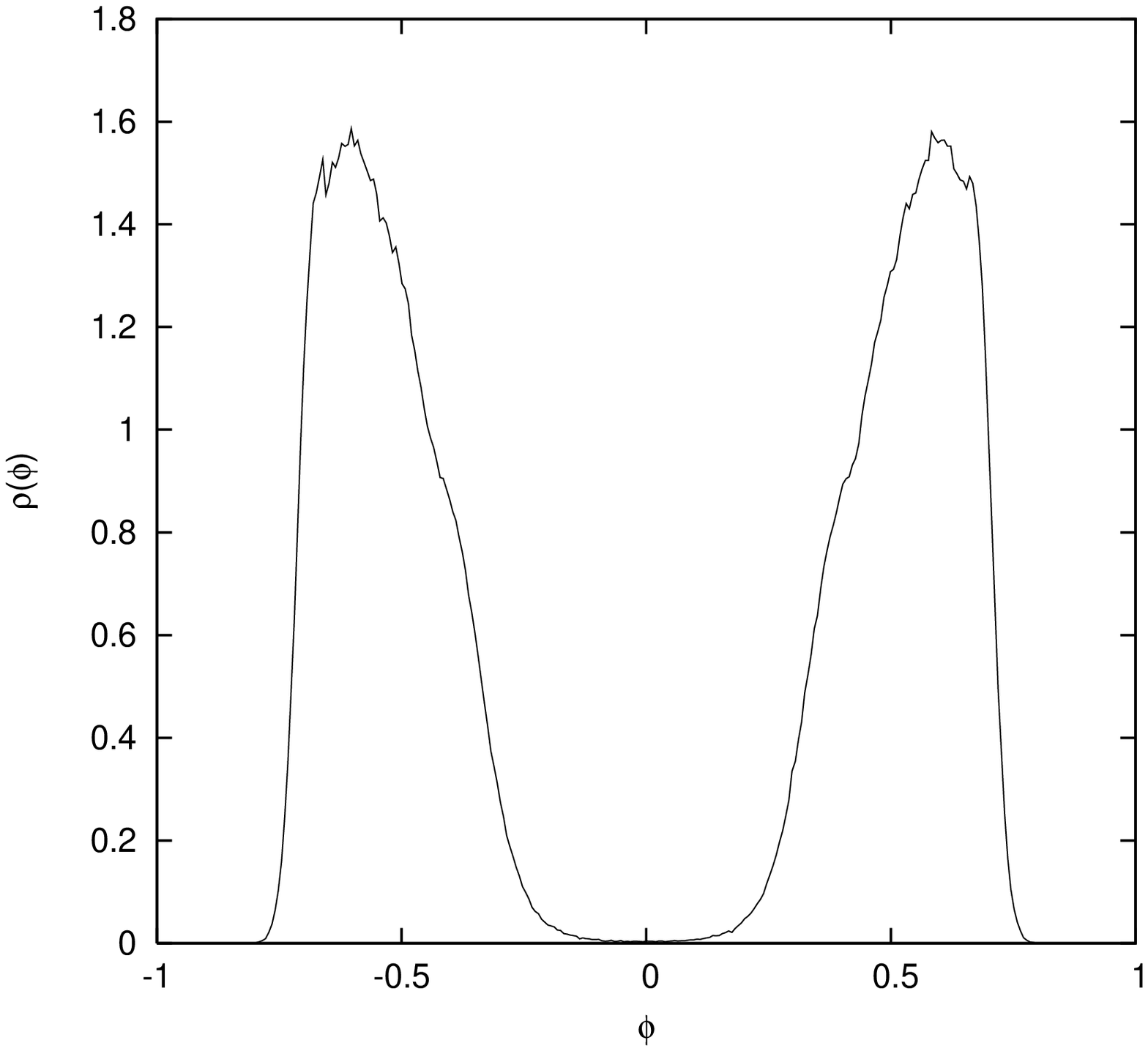}
\caption{When $\mu^2$ reaches a critical value, $\rho(0)$
vanishes; the data shown here were obtained at $\mu^2=-114.75$.}
\label{numerical_transition_fig}
\end{minipage}\hfill
\begin{minipage}[t]{75mm} \centering
\includegraphics[width=75mm]{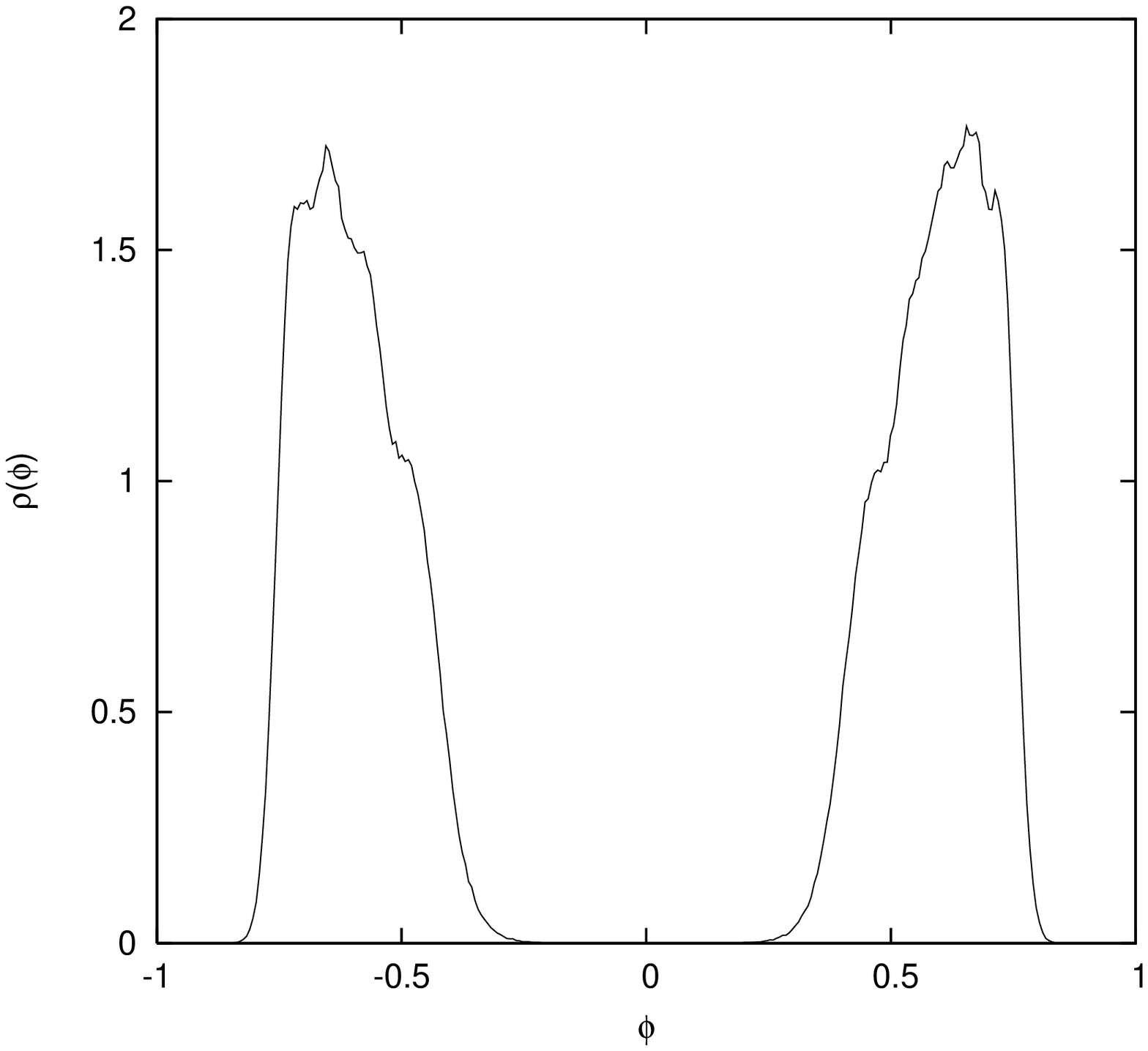}
\caption{The simulation results below the transition point (the
plot refers to data at $\mu^2=-135$) display a pattern compatible
with the ``two-cut phase'' \dots} \label{numerical_twocuts_fig}
\end{minipage}}
\vspace{-3mm}
\end{figure}

\begin{figure}[t]
\centerline{\begin{minipage}[t]{75mm} \centering
\includegraphics[width=75mm]{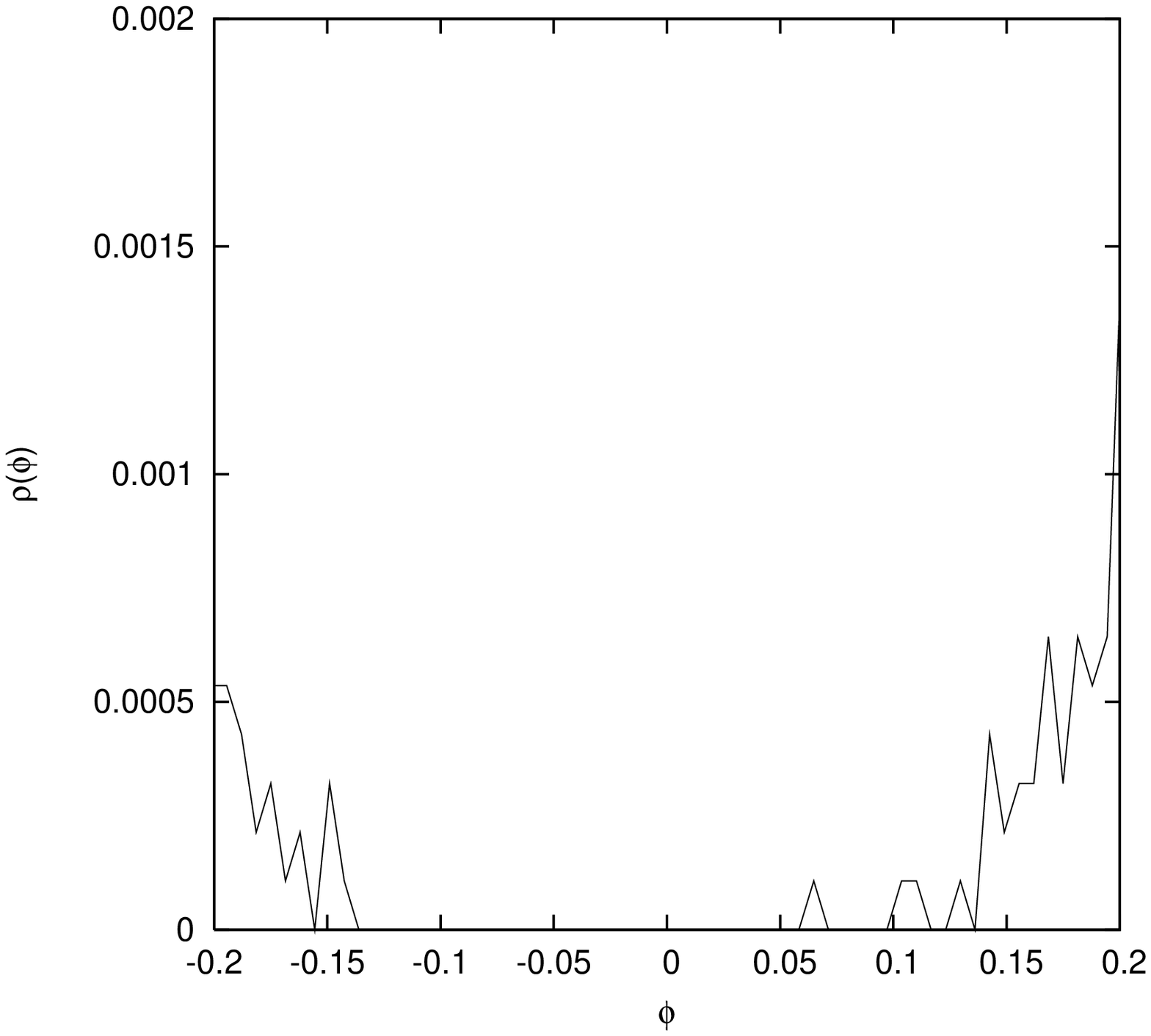}
\caption{\dots and a zero-density gap is observed in a f\/inite
interval around $\phi=0$.} \label{numerical_zoomed_twocuts_fig}
\end{minipage}\hfill
\begin{minipage}[t]{75mm} \centering
\includegraphics[width=75mm]{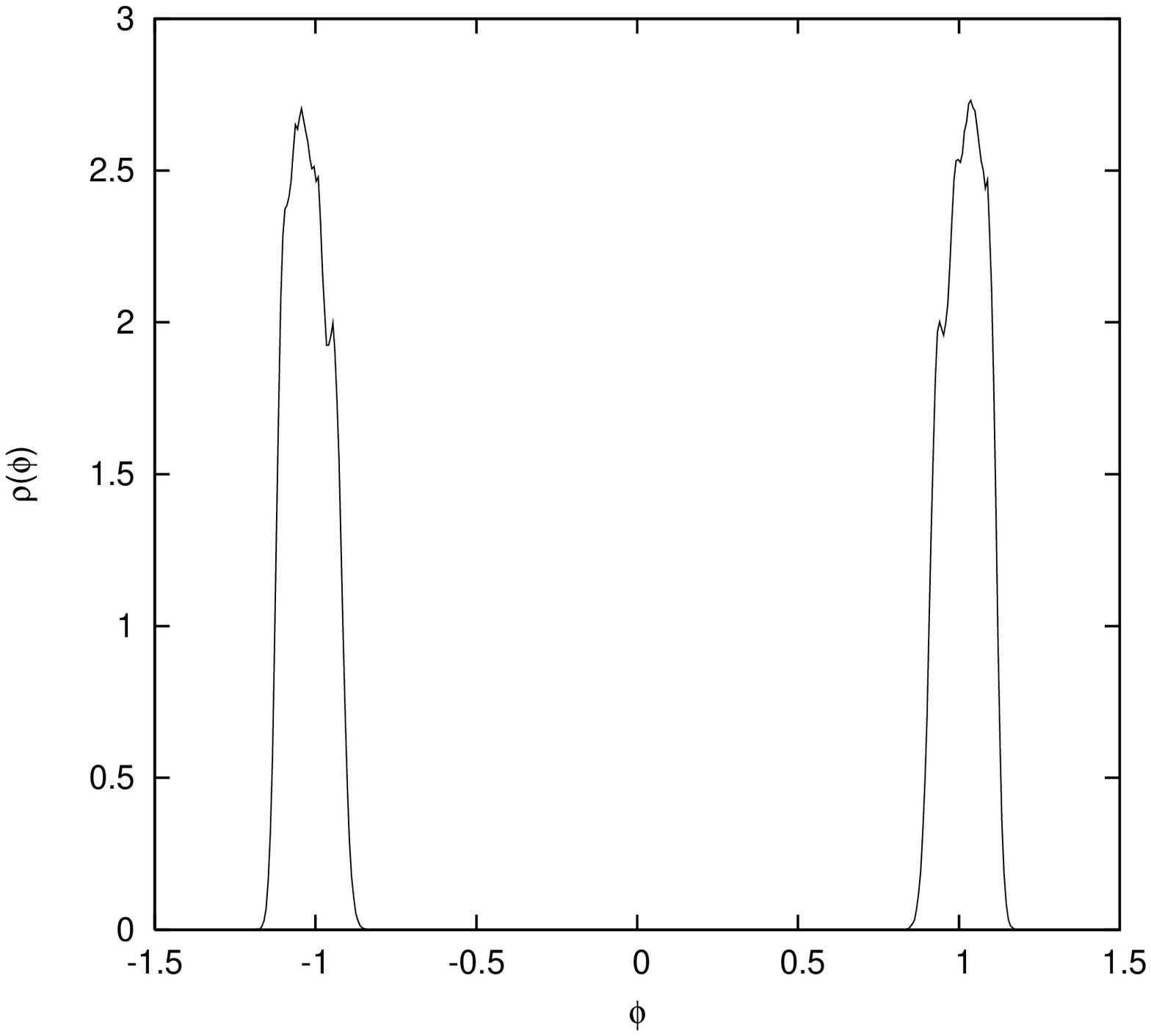}
\caption{For $\mu^2$ values deeper in the ``two-cut'' regime
($\mu^2=-351$), the distribution gets more localised around two
narrow peaks, and a zoom to the origin of this f\/igure gives
stronger evidence of vanishing eigenvalue density in a (larger)
neighbourhood of $\phi=0$: a large range exists, in which
\emph{no} eigenvalues are observed.}
\label{numerical_far_twocuts_fig}
\end{minipage}}
\vspace{-3mm}
\end{figure}

As an example, in Figs.~\ref{numerical_onecut_fig}
to~\ref{numerical_far_twocuts_fig} we display the plots obtained
from (low-statistics) independent data produced by test runs.
These simulations correspond to $N=9$, holding the
``non-commutativity parameter'' $\frac{2R^2}{N}=\frac{2}{9}$
f\/ixed and for constant $g=972$, with $\mu^2$ taking four
dif\/ferent, decreasing values. Note that the eigenvalues are not
rescaled to the $[-1,1]$ range, therefore the normalisation for
the $\phi$ axis is dif\/ferent with respect
to~\cite{Steinacker:2005wj} and to Figs.~\ref{semicircle_fig}
to~\ref{twocuts_fig}. Also, the errorbars and complete details of
the statistical analysis are not displayed for these preliminary
data -- they will be presented at the moment of the publication of
the f\/inal results.

The eigenvalue distribution plotted in
Fig.~\ref{numerical_onecut_fig} is obtained for $\mu^2=0$: its
features appear to be compatible with the ``one-cut'' prof\/ile
expected above the transition point (see Fig.~\ref{onecut_fig}).

For smaller values of $\mu^2$
(Fig.~\ref{numerical_almost_transition_fig}), the eigenvalue
density around $\phi=0$ decreases, and appears to eventually
vanish at a particular critical value:
Fig.~\ref{numerical_transition_fig} shows the distribution
prof\/ile approximately at the transition point\footnote{It should
be remarked that, in order to identify the transition properly,
one should perform a study of the scaling properties, when the
parameters are varied; this will be addressed elsewhere. Here, we
shall not make any quantitative claims about the precise location
of the critical lines. Still, the present discussion is (at least
qualitatively) consistent with the expected theoretical pattern,
and it provides the preliminary information necessary to address
the complete study.}.

Below the critical value, a ``two-cut'' pattern
emerges: the  numerical results shown in
Fig.~\ref{numerical_twocuts_fig} seem to conf\/irm that
$\rho(\phi)$ becomes peaked around two (symmetric) values, and,
more important, that a gap of zero probability opens up for
eigenvalues in a neighbourhood of zero. This is more clearly
visible in Fig.~\ref{numerical_zoomed_twocuts_fig}, which is a
zoom around the origin of the previous plot. Note, however, that
in this case we are on the borderline of the statistical precision
allowed by our sample, and the scale of the vertical axis is
already comparable with the minimal resolution of the data.

In order to obtain more reliable information and see more
precisely whether the eigenvalue distribution is indeed vanishing
in this regime, it is instructive to extend the analysis to
further, lower $\mu^2$ values. The result is displayed in
Fig.~\ref{numerical_far_twocuts_fig}, which conf\/irms the pattern
observed in Fig.~\ref{numerical_twocuts_fig}, showing a
distribution which is more strongly localised around two sharp
peaks; a zoom to the origin shows -- at least within the precision
of the data -- an exactly vanishing eigenvalue density in a larger
neighbourhood of $\phi=0$.

Not surprisingly, the general qualitative features emerging from
this picture are also shared by the three-dimensional model
discussed in~\cite{Medina:2005su}, in which two non-commutative
directions are combined with a (compactif\/ied) commutative
``time'', and the regularised model is described by the action in
equation~(\ref{julietamodel}). Indeed, also in that case, a phase
was observed, which is unknown in the model in ordinary
commutative space, and the dif\/ferent limits for $R$ and $N$
reproduce dif\/ferent physical theories.

\section{Conclusions}\label{conclusionsect}

To summarise the discussion presented above, we can conclude that
the fuzzy space formulation encodes a very rich structure,
describing dif\/ferent physical models.

On one hand, since the existence of ``anomalous'' ef\/fects, which
have no commutative counterpart, shows that the naive formulation
of the model would not reproduce the expected commutative theory,
it would be interesting to address a detailed study of some
``improved'' version of the matrix model, which may yield the
correct commutative limit.

On the other hand, the use of fuzzy spaces as a regularisation
scheme for intrinsically non-commutative models appears to be one
of the most powerful tools to investigate their non-perturbative
features, both analytically and numerically.

After reviewing the main problems in fuzzy models, we presented a
preliminary sample of new numerical results, which are in
agreement with the pattern predicted by the theory for the phase
transition between two dif\/ferent regimes in the scalar model on
the fuzzy sphere. In particular, we showed evidence supporting the
picture of a phase transition which can be investigated looking at
the topological properties of the matrix eigenvalue distribution.

Although in this paper we focused our attention onto some
specif\/ic issues, the problems related to fuzzy spaces are among
the main research interests of many dif\/ferent groups worldwide;
therefore we apologise with the authors whose results have not
been mentioned or properly discussed here. A lot of work has been
done, and a lot more is currently in progress in this exciting
research f\/ield -- a f\/ield which, in view of the present status
of the research, looks very promising.

\subsection*{Acknowledgements}

The author thanks the organisers of the O'Raifeartaigh Symposium
on Non-Perturbative  and Symmetry Methods in Field Theory (June
22--24, 2006, Budapest, Hungary) for the stimulating and really
enjoyable atmosphere at the conference, as well as
A.P.~Balachandran, W.~Bietenholz, B.P.~Dolan, K.S.~Gupta,
D.~O'Connor and H.~Steinacker for enlightening discussions. This
work is supported by Enterprise Ireland under the Basic Research
Programme.

\LastPageEnding

\end{document}